\documentstyle{article}

         \def\ba{\begin{array}}
         \def\ea{\end{array}}
         \def\be{\begin{equation}}
         \def\ee{\end{equation}}

         \def\D{{\Delta }}
         \def\P{{\Phi}}
         \def\p{{\phi}}
         \def\si{{\psi}}
         \def\d{{\partial}}
         \def\R{{\bf R}}
         \def\be{\begin{equation}}
         \def\ee{\end{equation}}
         \def\bea{\begin{eqnarray}}
         \def\eea{\end{eqnarray}}
         \def\ba{\begin{array}}
         \def\ea{\end{array}}

         \def\D{\Delta}

\def\I{\rm {I\kern-.3em I}}
\def\C{\rm {I\kern-.520em C}}
\def\R{\rm {I\kern-.3em R}}
\def\CZ{\rm {Z\kern-.4em Z}}

\def\unit{\rm {1\kern-.4em 1}}

\begin{document}
\begin{titlepage}
\vspace{-10mm}
%%%%%%%%%%%%%%%%%%%%  The title page   %%%%%%%%%%%%%%%%%%%%%%%%%%%%%%%%%%%%%
\hfill{}

%\vbox{
%    \halign{#\hfil         \cr
%            hep-th/9611009 \cr\noalign{\vskip -0.5cm}
%            IPM 96-        \cr\noalign{\vskip -0.5cm}
%            TUDP 96      \cr\noalign{\vskip -0.5cm}
%            IASBS 96     \cr\noalign{\vskip -0.5cm}
%           } % end of \halign
%      }  % end of \vbox
\vskip 30 mm
\centerline{ \Large \bf A pseudo--conformal representation of the Virasoro 
           algebra}
\vskip 10 mm
\centerline{ A. Aghamohammadi$^{1,2,*}$, M. Alimohammadi$^{1,3}$, 
            \& M. Khorrami$^{1,3,4}$ }

\vskip 10 mm
{\it
  \leftline{ $^1$ Institute for Studies in Theoretical Physics and
            Mathematics, P.O.Box 5531, Tehran 19395, Iran}
\vskip -.4mm
  \leftline{ $^2$ Department of Physics, Alzahra University,
             Tehran 19834, Iran. }
\vskip -.4mm  
  \leftline{ $^3$ Department of Physics, Tehran University,  
             North-Kargar Ave. Tehran, Iran. }
\vskip -.4mm  
  \leftline{ $^4$ Institute for Advanced Studies in Basic Sciences,
             P.O.Box 159, Gava Zang, Zanjan 45195, Iran. }
\vskip -.4mm  
  \leftline{ $^*$ e-mail: mohamadi@rose.ipm.ac.ir}
  }
\vskip 10mm
\begin{abstract}
Generalizing the concept of primary fields, we find a new representation of 
the Virasoro algebra, which we call it a pseudo--conformal representation. 
In special cases, this representation reduces to  
ordinary-- or logarithmic--conformal field theory. There are, however, 
other cases in which the Green functions differ from those of ordinary-- or 
logarithmic--conformal field theories. This representation is parametrized 
by two matrices. We classify these two matrices, and calculate some of the 
correlators for a simple example.
\end{abstract}
\vskip 10 mm
\end{titlepage}
%%%%%%%%%%%%%%%%%%%%  The body of the paper            %%%%%%%%%%%%%%%%%%%%%

In an ordinary conformal field theory primary fields are the highest weights
of the representations of the Virasoro algebra. A primary field
$\p (w,\bar w)$ can be defined through its operator product expansion with
the stress-energy tensor $T(z)$ ( and $\bar T (\bar z)$ ) or equivalently
through its commutation relations with the Laurent expansion coefficients
of $T$; $L_n$'s \cite{BPZ}:
\be \label{1}
[L_n,\p_i(z)]=z^{n+1}\d_z\p_i+(n+1)z^n\D_i \p_i,
\ee                               
where $\D_i$ is the conformal weight of $\p_i$.
One can regard $\D_i$'s as the diagonal elements of a diagonal matrix $D$,
\be \label{2}
[L_n,\p_i(z)]=z^{n+1}\d_z\p_i+(n+1)z^nD_{ij} \p_j.
\ee
One can however, extend the above relation for any matrix $D$, which is not
necessarily diagonal. This new representation of $L_n$  also satisfies the
Virasoro  algebra for any arbitrary matrix $D$ \cite{RAK}. By a suitable 
change of basis, one can make $D$ diagonal or Jordanian. If it is 
diagonalizable, the field theory is nothing but the ordinary conformal 
field theory (CFT).
Otherwise it should be in the Jordanian form. The latter case is the
logarithmic conformal field theory (LCFT) \cite{Gu,CK,RAK}. In the 
simplest case, the Jordanian block is two dimensional and the relation 
(\ref{2}) for the two fields $\p$ and $\si$, becomes
\be \ba{ll}
&[L_n,\p(z)]=z^{n+1}\d_z\p+(n+1)z^n\D \p \cr
&[L_n,\si(z)]=z^{n+1}\d_z\si+(n+1)z^n\D \si+(n+1)z^n \p .\ea
\ee
The field $\p$ is  an ordinary primary field, and the field $\si$
is called a quasi-primary or logarithmic field and they transform in the
following way:
\be \ba{ll}
&\p(z)\to ({\d f^{-1}\over \d z})^{\D}\p(f^{-1}(z))\cr
&\si(z)\to ({\d f^{-1}\over \d z})^{\D}[ \si(f^{-1}(z))+
\log ({\d f^{-1}(z)\over \d z}) \p(f^{-1}(z)]\ea
\ee
The two-point functions of these fields has been obtained in \cite{Gu,CK}. 
It has been shown in
[2]  that any $n-$point function ( for $n>2$ ) containing the field $\si$
can be obtained through the $n-$point function containing the field $\p$
instead of $\si$. 

Now the natural question which may arise is that,
``is it possible to generalize (2) such that $L_n$'s are still a
representation of Virasoro algebra?''
To investigate this question, we consider the following generalization of 
equation (2) 
\be \label{5}
[L_n,\p_i(z)]=z^{n+1}B_{ij}\d_z\p_j+(n+1)z^nA_{ij} \p_j +C_i,
\ee
and impose the condition that $L_n$'s satisfy the Virasoro algebra:
\be
[[L_n,L_m],\p_i ]=(n-m)[L_{m+n},\p_i ].
\ee
Now it is easy to see that the generalization (5) satisfies the Virasoro 
algebra provided that the matrices $A, B$ and $C$ satisfy the following 
relations:
\be B^2+BA-AB=B\ee
\be BA=A\ee
\be C=0\ee
Now we  try to classify the solutions of $A$ and $B$. The trivial solution is
$B=1$, which is nothing but CFT, when $A$ is diagonizable, and LCFT, when
$A$ is not diagonizable. If $A$ is an invertible matrix the only solution
for $B$ is $B=1$. However for any CFT which contains identity or any other
field with zero conformal weight, $A$ is not invertible, and there exists a
corresponding new theory with $B\ne 1$ for which $L_n$'s satisfy the
Virasoro algebra. This is obviously not a CFT any more, as the action 
of any diffeomorphism on a field contains a term $-\xi\cdot\partial\phi$, 
where $\xi$ is the generator of the diffeomorphism. This corresponds to 
$B=1$. For this reason, we call this representation a {\it pseudo--conformal} 
representation.
Using (7,8) we have
\be
(B-A)(B-1)=0.
\ee
Multiplying both sides of the above equation from the left by $B$, and using
(8), leads to
\be
(B-1)^2B=0.
\ee
This means that the eigenvalues of $B$ is equal to one or zero. So one can
take the matrix $B$ in the block diagonal form, where the blocks should be 
one of the following cases:

\noindent i) zero matrix

\noindent ii) identity matrix

\noindent iii) two dimensional Jordanian blocks
$\pmatrix{ 1 & 1 \cr 0 & 1 \cr }.$

We choose a basis in which the matrices $A$ and $B$ are in the following
form:
\be A=\pmatrix{A_1&\vert&A_2\cr\line(1,0){10}&\line(1,0){10}&\line(1,0){10}
\cr A_3&
\vert&A_4\cr}\qquad B=\pmatrix{B_1&\vert&0\cr--&-\vert -&-- \cr 0&\vert&0}
\ee
Using (8) and (10) it can be shown that $A_2=A_3=A_4=0$
and 
\be B_1A_1=A_1.
\ee
Now  the matrix $B_1$ can be written in the following form:
\be
B_1=\pmatrix{1&1\vert &0\cr \line(1,0){5} 0\line(1,0){5}&
\line(1,0){5}1\vert \line(1,0){5}&\line(1,0){5}0\line(1,0){5}
\cr 0&0\vert &1\cr},
\ee
where $1$ stands for the identity matrix and the dimension of the first
block should be even ($2k$). So
\be \label{11}
B_1-1=\sum^k_{i=1}e_{i,i+k},
\ee
where $(e_{ij})_{kl}=\delta_{ik}\delta_{jl}$ which together with (13) yields:
\be  \label{15}
(A_1)_{m+k,n}\Theta_{mk}=0 \qquad {\rm where}\qquad \Theta_{mk}=
\cases{1&$m\leq k$\cr 0&$m>k$\cr},
\ee
So all the elements of the lines $k+1$ to $2k$ of the matrix $A_1$ 
should be zero. Now it is easy to show that 
\be \label{13}
(A_1-1)(B_1-1)=0,
\ee
where we have used 
\be
(B_1-1)^2=0.
\ee
Substituting equation (15) in (17) results:
\be 
(A_1-1)_{m,n-k}=0,\hskip 2cm k<n\leq 2k.
\ee
and combining  (16) and (19), the matrix $A$ takes the following form
\be
A=\pmatrix{1&A'_1&A'_2&0\cr 0&0&0&0\cr 0&A'_3&A'_4&0\cr 0&0&0&0\cr},
\ee
Finally we use a similarity transformation which does not change $B$, to put 
the matrix $A$ in a simpler form:
\be \label{21}
A=\pmatrix{1&0&A'_2&0\cr 0&0&0&0\cr 0&A'_3&A'_4&0\cr 0&0&0&0\cr}
\qquad B=\pmatrix{1&1&0&0\cr 0&1&0&0\cr 0&0&1&0\cr 0&0&0&0\cr},
\ee
Now we are at the point to calculate the correlation functions of the field
$\p_i$, which have the Virasoro symmetry,i.e. 
\be
<[L_n,\p_i\p_j\cdots ]>=0  \qquad  n=0,\pm 1.
\ee
In general $A'_2,A'_3$ and $A'_4$ are arbitrary matrices. As an example we 
calculate two- and three-point functions in a simple case of (21):
\be
A=\pmatrix{1&0\cr 0&0\cr}\qquad B=\pmatrix{1&1\cr 0&1\cr},
\ee
where $1$ stands for identity matrix. If we write the fields in a coloumn 
matrix:
\be
\P =\pmatrix{(\p_i^1)\cr (\p_i^2)}
\ee
the relation (5) becomes:
\be \label{22}
[L_n,\p_i^1]=z^{n+1}\d_z(\p_i^1+\p_i^2)+(n+1)z^n\p_i^1
\ee
\be \label{23}
[L_n,\p_i^2]=z^{n+1}\d_z\p_i^2 
\ee
Noting (26) we see that $\p_i^2$'s are like ordinary primary fields, 
with zero conformal weight of a CFT, so 
\be
<\p_i^2(z)\p_j^2(w)>=c.
\ee
To calculate the other two-point functions, we use (25) and(26),
which leads to 
\be
<\p_i^1(z)\p_j^2(w)>=0
\ee
\be
<\p_i^1(z)\p_j^1(w)>={d\over (z-w)^2}
\ee
Surprisingly, the above two-point functions are the same as the two-point 
functions of a CFT, in which conformal weights of the fields are zero and 
one. Now consider the three-point functions. The simplest case is 
$<\p_i^2(z_1)\p_j^2(z_2)\p_k^2(z_3)>$. With a reason similar to that of 
the two-point functions, it is equal to:
\be
<\p_i^2(z_1)\p_j^2(z_2)\p_k^2(z_3)>=c_{ijk}
\ee
Using (25) and (26), one can also obtain 
\be
<\p_i^2(z_1)\p_j^2(z_2)\p_k^1(z_3)>={\alpha_{ijk} (z_1-z_2)\over 
(z_1-z_3)(z_3-z_2)}
\ee
where $c_{ijk} $ and $\alpha_{ijk}$ are some constants which depends on 
$i,j$ and $k$.
This is also similar to three-point function of the fields of weight zero 
and one in an ordinary CFT. The three-point function
$<\p_i^2(z_1)\p_j^1(z_2)\p_k^1(z_3)>$ can also be calculated, which results
\be
<\p_i^2(z_1)\p_j^1(z_2)\p_k^1(z_3)>={\alpha_{ijk} z_2+\alpha_{ikj}z_3
+d_{ijk}\over (z_2-z_3)^2}
\ee
where $d_{ijk}$ is a constant. This three-point function is completely 
different from the conformal case, and its dependence on coordinates is not 
only through their differences ( as in CFT's ).
This behaviour results from existance of nontrivial $B$ matrix in our 
algebra. The correlator $<\p_i^1(z_1)\p_j^1(z_2)\p_k^1(z_3)>$ can be also 
calculated, and it is seen that it is similar to that of ordinary CFT.

Using the above symmetry alone, one cannot obtain $n$-point functions with 
$n>3$, because there are only three first order partial differential 
equations for a function of more than three variables. This is as the case 
of ordinary  CFT. One can, however, restrict the function 
to an $n-3$ variable function, just as in the case of ordinary CFT, and 
this restriction differs from that of an ordinary CFT.

{\bf Acknowledgement} M. Alimohammadi and M. Khorrami wish to acknowledge 
the research vice-chancellor of Tehran University for partial support of 
this work.
%%%%%%%%%%%%%%%%%%%%%%%%%%%%%%%%%%%

\end{document}